\documentclass[submission,copyright,creativecommons]{eptcs}
\providecommand{\event}{ACL2 2022} 
\usepackage{underscore}           

\newcommand{\codestyle}[1]{\fontfamily{lmtt}\bfseries\selectfont #1}
\DeclareTextFontCommand{\code}{\codestyle}

\newcommand{\eg}{\emph{e.g.}}
\newcommand{\etal}{\emph{et al.}}

\newcommand{\stobj}{stobj}
\newcommand{\computefn}{\code{acl2s-compute}}
\newcommand{\queryfn}{\code{acl2s-query}}
\newcommand{\eventfn}{\code{acl2s-event}}
\newcommand{\simplequeryfn}{\code{acl2s-query-simple}}

\newcommand{\seexdoc}[1]{(X:\href{https://www.cs.utexas.edu/users/moore/acl2/manuals/current/manual/?topic=ACL2____#1}{\MakeLowercase{#1}})}

\newcommand{\Pcomment}[1]{}
\usepackage{textcomp}
\usepackage{upquote}
\usepackage{amsmath}
\usepackage[noend]{algpseudocode}
\usepackage{acl2}

\title{ACL2s Systems Programming}
\author{Andrew T. Walter \& Panagiotis Manolios
\institute{Northeastern University}\\
\email{walter.a@northeastern.edu, pete@ccs.neu.edu}}

\def\authorrunning{Walter \& Manolios}
\def\titlerunning{ACL2s Systems Programming}
\begin{document}
\maketitle

\begin{abstract}
  ACL2 provides a systems programming capability that allows one to
  write code that uses and extends ACL2 inside of ACL2. However, for
  soundness reasons, ACL2 bars the unrestricted use of certain kinds
  of programming constructs, like destructive updates, higher-order
  functions, eval, and arbitrary macros.  We devised a methodology for
  writing code in Common Lisp that allows one to access ACL2, ACL2s,
  and Common Lisp functionality in a unified way. We arrived at this
  methodology in the process of developing the ACL2 Sedan (ACL2s) and
  using it as a key component in formal-methods-enabled projects
  relating to gamified verification, education, proof checking,
  interfacing with external theorem provers and security. The
  methodology includes a library for performing ACL2 queries from
  Common Lisp, as well as guidelines and utilities that help address
  common needs. We call this methodology ``ACL2s systems
  programming,'' to distinguish it from ACL2 systems programming. We
  show how our methodology makes it possible to easily develop tools
  that interface with ACL2 and ACL2s, and describe our experience
  using it in our research.
\end{abstract}

\section{Introduction}

Since its inception, users of ACL2~\cite{acl2-car, acl2-acs, acl2, acl2-web}
have developed many tools that extend ACL2 or use ACL2 as an integral
component. Extensions include improved termination
analysis~\cite{ccg}, the use of external
solvers~\cite{manolios-framework-verifying-pipelined,
  manolios-framework-verifying-pipelined-journal, reeber-sulfa,
  peng-smtlink, smtlink2}, support for bit-blasting
proofs~\cite{manolios-framework-verifying-pipelined,
  manolios-framework-verifying-pipelined-journal, swords-bit-blasting}
and automatic counterexample
generation~\cite{chamarthi-integrating-testing}. Systems that are
built on top of ACL2 include
ACL2r~\cite{gamboa-acl2-nonstandard-analysis} and the ACL2 Sedan
(ACL2s)~\cite{dillinger-acl2-sedan}. An underlying philosophy that
seems to have guided the development of these tools is that as much of
their functionality as possible should be implemented in ACL2, using
ACL2's systems programming support. The ACL2 developers and the
community have developed capabilities to support the development of
tools using this philosophy, including \code{make-event}, trust tags
and the ``hacking'' books~\cite{dillinger-hacking}. In this paper, we
advocate for a different philosophy, which we call ``ACL2s systems
programming,'' that prioritizes the seamless integration of ACL2 and
Common Lisp code. By jettisoning the requirement that as much of a
tool's code be written inside of ``core ACL2''---that is, without
installing trust tags or dropping out of ACL2's REPL---one can write
code that uses functionality that ACL2 limits or does not support,
like higher-order functions, destructive operations, and external
libraries. The cost of doing so is the loss of guarantees and checking
that ACL2 provides for code written in ACL2. In some applications
these guarantees are worth the cost of heeding ACL2's limits on
functionality, but we have found ourselves developing tools where the
use of the ACL2 systems programming approach leads to brittle code
that is hard to write, understand, modify and update.

Motivated by our experience, we have developed a library that provides
an easy-to-use interface (the ``ACL2s interface'') for performing ACL2
queries and gathering their results from ``raw Lisp''---that is, from
a Common Lisp process that has ACL2 loaded. This library provides
additional functionality that is nontrivial to implement, such as the
ability to capture all ACL2 printed output. Based on our experiences using
this library, we have also developed a set of guidelines and general
advice around how best to write Common Lisp code that communicates
with ACL2 via our library.  By using our library and adhering to our
guidelines, programmers can write ``systems ACL2s'' code that uses the
full feature set of Common Lisp and also interacts with ACL2, while
clearly denoting which code is directly interacting with the theorem
prover. We note that we call our methodology ``ACL2s systems
programming'' not because it is limited to use with the ACL2 Sedan
system, but rather to help disambiguate our work from the existing
ACL2 systems programming utilities and because the methodology was
developed in the context of building tools based on ACL2s.

To understand how the ability to write Common Lisp that calls ACL2 is
useful, let us consider some tasks that a tool using ACL2 might need to
perform:
\begin{itemize}
\item \textbf{Domain-Specific Language Support}: Consider developing a
  domain-specific language that includes verification capabilities
  provided by ACL2. A first step is to define a parser, which can be
  done in ACL2, but architecting a parser in such a way that it is
  able to provide high-quality error messages is difficult and
  tedious.  By directly using Common Lisp, one can use existing
  parsing libraries like ESRAP~\cite{esrap} to take care of the
  details of error handling while also greatly simplifying the process
  of writing a parser.  To use these libraries, one may be required to
  make use of Common Lisp features that ACL2 does not support, for
  example the condition system, CLOS, or higher-order functions. There
  is no reason why one could not write similar libraries in ACL2, but
  to our knowledge no such libraries exist yet.
  Alternatively, one can interface with an external system, like
  Xtext~\cite{xtext}, that manages parsing and error messages and
  provides support for developing tools that run in browsers. ACL2s
  systems programming can be used to run sophisticated validation
  checks in Xtext and for verifying code written in the DSL.
\item \textbf{Generating and Running ACL2 Queries}: Some queries are
  difficult to generate using ACL2 code. For example, accessing
  the user's current package using ACL2 requires having access to
  the \code{state} \stobj. This means that any function that might
  need to know the user's current package
  must take in \code{state}. This means that the function's callers
  must also take in \code{state}, and its callers' callers, and so
  on. Thus, giving an existing function access to \code{state} so that
  it can determine the user's current package might require
  significant changes to several other functions. If you do need
  access to \code{state} to build a query, you cannot implement the
  query-building using macros. Instead, you must use functions like
  \code{make-event} to build and execute the query. Using
  \code{make-event} introduces more considerations, as there are
  limitations on what forms may appear inside of \code{make-event}
  calls, as well as limitations on where \code{make-event} calls can
  be placed. Additionally, one might find themselves nesting
  \code{make-event}s when building large queries using smaller
  queries, making the code more difficult to debug and harder to
  generate useful error messages for. Building a system using
  \code{make-event}, like the \code{defdata} functionality inside
  ACL2s~\cite{chamarthi-defdata}, can be worth it for the guarantees
  that ACL2 provides for core ACL2 code. However, when one does not
  intend to reason about the code that is generating ACL2 queries, the
  complexity that comes with writing such code in core ACL2 may
  outweigh the benefits of writing core ACL2.
\item \textbf{Handling Requests from a Web Client}: There are two
  approaches to enabling ACL2 to handle requests from a web client:
  either use another programming language to handle the details of
  receiving and responding to network requests and have that language
  communicate with an ACL2 subprocess, or have ACL2 directly handle
  networking. The latter could be done by using a library like
  Hunchentoot~\cite{hunchentoot} (which has a systems ACL2
  wrapper). The ACL2 Bridge~\cite{davis-acl2-bridge} is an example of
  an ACL2 system that takes the latter approach. The former has also
  been done before---see the ACL2 Sedan Eclipse
  plugin~\cite{dillinger-acl2-sedan}, which runs an ACL2 subprocess
  inside of a Java process. However, building a robust system that can
  reliably communicate with an ACL2 subprocess takes significant
  engineering effort---Dillinger \etal\ developed the ACL2 hacking
  community books~\cite{dillinger-hacking} during their work, since
  they needed to be able to redefine built-in ACL2 functions to ensure
  they could accurately classify and manage ACL2's output. To get the
  ACL2 Sedan working, they also needed to develop a Lisp parser inside
  of Java, as well as an enhanced history tracking system for ACL2.
\end{itemize}


As a concrete example, consider a current project that uses the ACL2
Sedan's (ACL2s') counterexample
generation~\cite{chamarthi-integrating-testing} and data
definition~\cite{chamarthi-defdata} functionality to generate members
of data types that are requested by another (unverified) system. For
this project, it is convenient to communicate with the other system
over TCP sockets using JSON encoding, so we want to use Common Lisp
libraries for doing so. Further, we would like to use the Common Lisp
condition system to gracefully handle errors that the Common Lisp
networking library might emit. We need to be able to generate certain
kinds of code that would be difficult to generate given ACL2's limits
on macros. Having access to the Common Lisp Object System would make
it easier to write certain parts of the system in a more easily
extensible fashion. Due to all of these considerations, it would take
a significant amount of effort to develop this project inside of ACL2,
and indeed doing so would require us to trust some raw Lisp code at
some point for interacting with TCP sockets. Since the ACL2s portion
of the project is going to be composed with an unverified system and
we thus cannot reason about the composite, the consistent logical
story that ACL2 can provide is not useful in this application and
trying to shoehorn the functionality we need in ACL2 leads to brittle
code that is hard to write, understand, modify and update.

The primary contribution of this paper is a description of the design
and implementation of an interface that allows for convenient calls
from Common Lisp into ACL2. We also provide a discussion of a
programming methodology that we have developed around the use of this
interface, and its use in several applications. The interface code is
freely available online at
\url{https://gitlab.com/acl2s/external-tool-support/interface}. We
primarily use SBCL-hosted ACL2 builds, and therefore we have high
confidence in the interface code's support for such systems. The
interface code certifies successfully on CCL-hosted ACL2 builds and
appears to work correctly, but we have not performed enough testing to
be confident in its behavior on such systems. We are more than happy
to implement support for ACL2 running on other Common Lisp
implementations upon request.

The remainder of this paper is organized as follows. We discuss the
three functions at the core of the ACL2s interface in
Section~\ref{sec:interface}. We provide a specification for these
functions in Section~\ref{sec:specification}. We describe our
experience using systems ACL2s and the interface in various
applications in Section~\ref{sec:applications}. Related work is
discussed in Section~\ref{sec:related} and we end with conclusions and
acknowledgments in Section~\ref{sec:conclusion}.

\section{The ACL2s Interface}
\label{sec:interface}

In our description of the ACL2s interface, we assume that the reader
is familiar with the concepts of ACL2 function signatures,
multiple-value returns, single-threaded objects (\stobj s), and IO
channels. The ACL2 XDOC documentation system provides explanations of
each of these topics; the reader is highly encouraged to refer to XDOC
to review any topics they may be unfamiliar with. Below, we refer
to XDOC topics using with the syntax \seexdoc{FOO} to refer to the
topic ``foo.''

ACL2 is designed in such a way that function definitions inside of the
ACL2 logic result in Common Lisp function definitions. That is, one
can define a function in the ACL2 REPL, exit out into raw Lisp, and
call that function from raw Lisp. We make extensive use of this
functionality in tools developed using the ACL2s systems programming
methodology. However, some functions (especially those that deal with
\stobj s) will either raise an error when called from raw Lisp or
behave incorrectly. This is a safety measure---the ACL2 REPL takes
care of several details (like setting readtables, handling function
guards, and ensuring that \stobj s are updated correctly) that
differentiate ACL2 evaluation from Common Lisp evaluation.  It is
possible to work around these restrictions, but doing so can be
difficult and error-prone, hence, for when ACL2 behavior is needed, we
developed an interface function that uses ACL2's built-in interface
for evaluating forms (\code{ld}). Running ACL2 code through \code{ld}
has the advantages just described, but comes with with a performance
cost.

We provide three raw-Lisp functions for interfacing with ACL2:
\computefn, \queryfn, and \eventfn. \computefn\ is the simplest of the
three, and it supports evaluating expressions that return a single
value and do not make use of \stobj s. \queryfn\ supports evaluating
expressions that return an error triple
\seexdoc{PROGRAMMING-WITH-STATE}, and is intended for queries that do
not involve calls to ACL2 event functions \seexdoc{EVENTS}. \eventfn\
supports evaluating expressions that return an error triple, and is
intended for queries that result in calls to ACL2 event functions. All
three of these functions are implemented by making a call to \code{ld}
with appropriate arguments. \code{ld} does not return the results of
the evaluated expressions, so we also insert code that extracts the
result of the given expression evaluation and stores it for access
after \code{ld} finishes.

The interface exposes three functions partially due to the way it
evolved over time. \computefn\ was designed to be a lightweight
function that only operates on non-multiple-value terms, \queryfn\ was
designed to operate on multiple-value terms that do not generate ACL2
events and \eventfn\ was designed to evaluate terms that contain ACL2
event functions and where the return value outside of the error flag
is considered unimportant.

\subsection*{Implementation}

To better understand the implementation of the ACL2s interface, let us
consider a simplified definition of \queryfn, \simplequeryfn, shown in
Figure~\ref{fig:acl2s-query-simple}. This function is missing much of
the functionality of \queryfn---for example, it does not have any
output control features and is missing some error handling. However,
it highlights the general structure of the ACL2s interface functions,
so we review it first.

\begin{figure}
\begin{verbatim}
;; this package imports only Common Lisp symbols
(in-package "ACL2S-INTERFACE-INTERNAL")
(defun acl2s-query-simple (q)
  (acl2::ld `((acl2::mv-let (erp val acl2::state)
                            ,q
                            (acl2::assign result (list erp val)))))
  (acl2::@ result))
\end{verbatim}
\caption{The definition of a simplified version of \queryfn}
\label{fig:acl2s-query-simple}
\end{figure}

\simplequeryfn, like \queryfn, is designed to evaluate ACL2
expressions that return ACL2 error triples and do not generate ACL2
events. As previously stated, \code{ld} does not return the result of
evaluating the ACL2 expressions it is given. Thus, we must construct
an ACL2 expression to run inside \code{ld} that allows us to capture
both the error flag (\code{erp}) and the returned value (\code{val})
from the error triple that the given ACL2 expression evaluates to.
For this purpose, we use the \code{mv-let} form. Note that, since we
are in a package that does not import ACL2 symbols like \code{state},
we must explicitly specify that we are binding to the ACL2 package's
\code{state} variable, not to a new state variable. If we forget to
specify the package for that symbol, ACL2 will produce a helpful error
message when \simplequeryfn\ is called, informing us that the \stobj\
\code{state} appears in a position where an ``ordinary object'' is
expected. We note that neither \simplequeryfn\ nor \queryfn\ currently
perform any checking to ensure that the provided expression does not
modify the world---this is a topic of future work.

Once we have evaluated the given expression and bound the constituents
of the error triple it returned to variables, we just need to store
\code{erp} and \code{val} somewhere that we will be able to access after
\code{ld} finishes executing. We choose to save the values in the
\code{global-table} of ACL2's \code{state}. After the call to
\code{ld} has completed, we use ACL2's \code{@} command to access the
value we inserted into the \code{global-table}.

Note that this approach to capturing the return value(s) of an ACL2
expression requires that the user know the output arity of the
expression that they are evaluating. In the context of ACL2s systems
programming, we typically are either dealing with functions that
return a single value or functions that return an error triple, so
\computefn, \queryfn, and \eventfn\ are sufficient. When dealing with
arbitrary ACL2 functions, one may need to interact with functions that
have a variety of multiple-value signatures, possibly containing
user-defined \stobj s. In this context, it may be impractical to
define a new interface function for each return signature that one
wants to support. There do exist alternate ways of evaluating ACL2
expressions and capturing their output that---like
\code{trans-eval}---operate on expressions of arbitrary output
arity. Such methods that we have tried impose additional restrictions
on the kinds of expressions that can be evaluated---for example,
\code{trans-eval} does not fully support evaluating \code{make-event}
expressions when invoked from raw Lisp. Another potentially useful
function is \code{ev}, which \code{make-event} uses internally to
evaluate ACL2 forms. We decided against using \code{ev}, as it is an
internal ACL2 function (it does not have an XDOC topic, and it is
marked as untouchable) and thus we cannot depend on its interface
being stable. Additionally, \code{ev} is a lower-level function with a
more complex set of preconditions that it imposes on its inputs than
\code{ld}, meaning that our interface functions would need to
implement some of the checks that \code{ld} performs for us,
introducing more room for error.

To provide an interface that allows a user to evaluate an
arbitrary-output-arity ACL2 form, one might consider trying to
determine the output arity of the form before evaluating it, and then
generating the appropriate \code{mv-let} form for the form's output
arity. However, it is difficult to statically determine the output
arity of an ACL2 expression (as one might do to generate the
appropriate \code{mv} binding form), since the number of values
returned may depend on which branches are taken for conditionals in
the expression.

Now, let us take a look at a more complete definition of \queryfn,
shown in Figure~\ref{fig:acl2s-query-full}. This definition is
considerably more complex, as in addition to the core \simplequeryfn\
functionality it handles output control (both capturing and disabling
output). We discuss output control in more depth later, but first
we explain some of the other functions that are used in \queryfn:

\begin{itemize}
\item \code{get-prover-step-limit} accesses ACL2's defaults table to
  determine if the user has set a step limit, and if not will return
  the default ACL2 step limit.
\item \code{+COMMAND-RESULT-VAR+} is a constant holding the symbol
  that we are storing the result of command evaluation in.
\item \code{ld-options} calls \code{ld} with the given set of keyword
  arguments inside of \code{with-suppression}. \code{with-suppression}
  is an ACL2 macro that is used inside of \code{lp} (ACL2's REPL
  function) to disable many kinds of warnings and errors that the
  underlying Common Lisp implementation may emit; without this, one
  may experience package lock errors and redefinition warnings among
  other diagnostics when working with innocuous core ACL2 code.
\item \code{last-result} calls gets the value of the entry named
  \code{+COMMAND-RESULT-VAR+} in ACL2's globals table by calling
  \code{f-get-global}.
\item \code{remove-props} removes plist entries with the given set of
  names from a plist. This is necessary to avoid passing any keyword
  arguments that \code{ld} does not know how to handle to \code{ld}.
\end{itemize}

\begin{figure}
  \small
  \begin{verbatim}
(in-package "ACL2S-INTERFACE-INTERNAL")
(defun acl2s-query (q &rest args &key quiet capture-output
                    (prover-step-limit (get-prover-step-limit)) &allow-other-keys)
  (let ((turned-quiet-mode-on (and quiet (not *quiet-mode-state*))))
    (when turned-quiet-mode-on (quiet-mode-on))
    (get-captured-output) ;; clear captured output
    (let ((state acl2::*the-live-state*))
      (ld-options `((acl2::with-prover-step-limit
                     ,prover-step-limit
                     (acl2::mv-let
                       (erp val acl2::state)
                       ,q
                       (acl2::assign ,+COMMAND-RESULT-VAR+ (list erp val)))))
                  (append (remove-props args '(:quiet :capture-output 
                                               :prover-step-limit
                                               :standard-co :proofs-co
                                               :ld-error-action))
                          `(:standard-co ',(calculate-standard-co args state)
                            :proofs-co ',(calculate-proofs-co args state)
                            :ld-error-action :error)
                          (when *quiet-mode-state* LD-QUIET-FLAGS)))
      (cleanup-streams)
      (when turned-quiet-mode-on (quiet-mode-off))
      (last-result))))
  \end{verbatim}
  \normalsize
  \caption{The full definition of \queryfn}
  \label{fig:acl2s-query-full}
\end{figure}

\subsubsection*{Output Control}
Control of ACL2's output is difficult to achieve---there are several
options that can be provided to disable most output (setting
\code{standard-co} and \code{proofs-co}, changing the
\code{inhibit-output-list}, enabling \code{gag-mode}), and until
recently some output (in particular, comment window output) was
impossible to disable without raw Lisp (for reference, see ACL2 commit
\code{709d5a55}).  To this end, the interface contains an
enabled-by-default feature that overwrites \code{comment-window-co},
an ACL2 function that is called to get the output channel for
comment-window \seexdoc{CW} output.  By creating our own channels and
choosing which one to return as the comment-window output channel
based on settings that the user provides, we can disable, capture, or
pass-through any comment window output from ACL2. The ability to
capture ACL2 output is particularly useful when making ACL2 queries
that may print out information that is not accessible through return
values, for example the contents of error messages. We can use the
\code{standard-co} and \code{proofs-co} \code{ld}-specials to capture
or disable that input using our own channels as well.

An ACL2 channel is simply a symbol with particular properties set in
its property-list. The names of these properties are stored in the
ACL2 values \code{*open\--output\--channel\--type\--key*} and
\code{*open\--output\--channel\--key*}. The first property is associated
with the IO type of the channel, which for the purposes of printed
output is always going to be \code{:character}. The second property is
associated with a Common Lisp stream that backs the channel. Our
interface code creates channels directly, by generating symbols and
setting the appropriate property-list properties, rather than using
ACL2's built-in interface for creating output channels. This is
because ACL2's interface for creating channels does not allow the user
to specify a backing stream for a channel. An alternative approach
might be to use ACL2's interface for creating channels, and then
overwrite the created channel's backing stream property with our own
stream, but this approach seemed just as likely to cause problems as
creating channels directly.

Output can be disabled by redirecting output to a channel backed by an
empty broadcast stream. Any output written to such a stream is simply
dropped. Capturing output is done by using a channel backed by a
string-output stream. In situations where we want to capture output
and also print it as normal, we can output to our own channel backed
by a stream that broadcasts to both our string-output stream and the
backing stream for the channel that the original
\code{comment-window-co} function returns. As of right now, we create
and garbage-collect (\code{cleanup-streams} in the definition of
\queryfn) these broadcast streams every time our implementation of
\code{comment-window-co} is called, but to improve performance we may
cache such streams and intelligently clear the cache when any relevant
settings change.

Users of our interface can control whether output should be disabled
or not either by providing the \code{:quiet} keyword argument to any
of the three interface functions or by calling \code{quiet-mode-on} or
\code{quiet-mode-off}. When quiet-mode is on, we prevent as much
output as possible from being printed by ACL2. Independently of
quiet-mode, users can control whether output is captured by either
using the \code{:capture-output} keyword argument or by calling the
\code{capture-output-on} or \code{capture-output-off} functions. The
captured output is cleared at the beginning of calls to the three
interface functions, so the user is responsible for saving captured
output after each call if they wish to persist it. Calling
\code{get-captured-output} will return a string corresponding to the
output produced by the last ACL2 interface call, but will also clear
that output from the stream.

To improve performance, it is useful to disable as much ACL2 output as
possible, as even writing to an empty stream carries a performance
penalty. Some ACL2 books may provide their own interfaces for turning
on or off their output, so we provide an interface by which users can
specify which ACL2 queries to execute when turning on quiet-mode or
turning it off. We use this interface in our ACL2s-specific code that
is provided alongside the ACL2s-agnostic code, a subset of which is
shown in Figure~\ref{fig:acl2s-quiet-mode}. This code creates a global
variable to save the value of the setting at the moment that
quiet-mode is turned on, so that it can be reset to this value when
quiet-mode is turned off. Hooks are registered with names so that they
can be removed or redefined later as needed.

\begin{figure}
  \small
  \begin{verbatim}
(in-package "ACL2S-INTERFACE-INTERNAL")
(defvar *saved-verbosity-level* acl2s::(acl2s-defaults :get verbosity-level))
(add-quiet-mode-on-hook :acl2s-verbosity-level
  (lambda ()
    (setf *saved-verbosity-level* acl2s::(acl2s-defaults :get verbosity-level))
    'acl2s::((acl2s-defaults :set verbosity-level 0))))
(add-quiet-mode-off-hook :acl2s-verbosity-level
  (lambda ()
    `((acl2s::acl2s-defaults :set acl2s::verbosity-level ,*saved-verbosity-level*))))
  \end{verbatim}
  \normalsize
  \caption{Setting up the interface to call book-specific code when enabling or disabling quiet-mode}
  \label{fig:acl2s-quiet-mode}
\end{figure}

\computefn\ (Figure~\ref{fig:acl2s-compute-full}) is defined similarly
to \queryfn. A \code{mv-let} inside of the call to \code{ld} is not
necessary, since \computefn\ is designed to evaluate queries that only
return a single value. Errors during evaluation are detected by
capturing the output of \code{ld}, which is an error triple. Based on
whether \code{ld} reports that it was able to successfully evaluate all
of the forms it was given (in this case, it returns \code{:eof} in the
value position of its error triple), \computefn\ either sets the
global table entry named \code{+COMMAND-RESULT-VAR+} with a value
indicating that no error occurred and including the value produced by
the evaluation of the given query, or a value indicating that an error
occurred.

As stated previously, functions defined in ACL2 can often be run
directly from raw Lisp, especially if they do not involve \stobj
s. So, under what circumstances would one want to use \computefn? It
provides a few major benefits over directly calling a function:
\begin{itemize}
\item A higher level of safety---\code{ld} will typically check that
  any arguments in a function call satisfy that function's guards (if
  any), while running the function directly from raw Lisp will not.
\item Better error handling---\code{ld} will give an error if your
  function call is not a valid ACL2 term for any reason, for example
  if you try to use an unbound variable in the call.
\item Output handling functionality---our interface, plus \code{ld},
  provide greater control over the output that the call to the
  function may attempt to print.
\item raw Lisp restrictions---Certain functions, \eg, those
  that contain a call to \code{acl2-unwind-protect}, will signal an
  error when they are run outside of the ACL2 loop.
\end{itemize}

\begin{figure}
  \small
  \begin{verbatim}
(in-package "ACL2S-INTERFACE-INTERNAL")
(defun acl2s-compute (c &rest args &key quiet capture-output
                      &allow-other-keys)
  (let ((turned-quiet-mode-on (and quiet (not *quiet-mode-state*))))
    (when turned-quiet-mode-on (quiet-mode-on))
    (get-captured-output) ;; clear captured output
    (let ((state acl2::*the-live-state*))
      (acl2::mv-let (erp val state)
        (ld-options `((acl2::assign ,+COMMAND-RESULT-VAR+ ,c))
                     (append (remove-prop args :quiet :capture-output :ld-error-action)
                             `(:standard-co ',(calculate-standard-co args state)
                               :proofs-co ',(calculate-proofs-co args state)
                               :ld-error-action :error)
                              (when *quiet-mode-state* LD-QUIET-FLAGS)))
                     (if (equal val :eof)
                         (save-result `(list nil (@ ,+COMMAND-RESULT-VAR+)))
                         (save-result `(list t nil))))
      (cleanup-streams)
      (when turned-quiet-mode-on (quiet-mode-off))
      (last-result))))
\end{verbatim}
  \normalsize
  \caption{The full definition of \computefn}
  \label{fig:acl2s-compute-full}
\end{figure}

\eventfn\ (Figure~\ref{fig:acl2s-event-full}) is defined similarly to
\queryfn. Unlike both \computefn\ and \queryfn, the return value(s) of
the provided expression is not captured. Instead, \eventfn\ returns
the \code{erp} component of the \code{ld} call, which the caller of
\eventfn\ can use to determine if their expression was admitted by
ACL2. For the sake of consistency, a length-2 list is returned, where
the second element is always \code{nil}.

\begin{figure}
  \small
  \begin{verbatim}
(in-package "ACL2S-INTERFACE-INTERNAL")
(defun acl2s-event (e &rest args &key quiet capture-output
                    (prover-step-limit (get-prover-step-limit))
                    &allow-other-keys)
  (let ((turned-quiet-mode-on (and quiet (not *quiet-mode-state*))))
    (when turned-quiet-mode-on (quiet-mode-on))
    (get-captured-output) ;; clear captured output
    (let ((state acl2::*the-live-state*))
      (acl2::mv-let (erp val state)
         (ld-options `((acl2::with-prover-step-limit
                         ,prover-step-limit
                         ,e))
           (append (remove-props args '(:quiet :capture-output :prover-step-limit
                                        :ld-error-action))
                   `(:standard-co ',(calculate-standard-co args state)
                     :proofs-co ',(calculate-proofs-co args state)
                     :ld-error-action :error)
                   (when *quiet-mode-state* LD-QUIET-FLAGS)))
         (progn
           (setf erp (not (equal val :eof)))
           (cleanup-streams)
           (when turned-quiet-mode-on (quiet-mode-off))
           (list erp nil))))))
\end{verbatim}
  \normalsize
  \caption{The full definition of \eventfn}
  \label{fig:acl2s-event-full}
\end{figure}

\subsection*{Guidelines for Use}

As we have built and used the interface, we have developed a set of
guidelines that help avoid common bugs when using the interface.

\begin{itemize}
\item \textbf{Operate in a separate package} - Though we do not want
  to reason about the code we are writing, we still would like to
  avoid accidentally overwriting any of ACL2's internals. For this
  reason, it is safest to operate in a package outside of \code{ACL2}
  or any other package defined in ACL2. Note that when operating in a
  separate package, one must be careful when constructing queries that
  contain symbols in the ACL2 package. To reduce the burden of
  explicitly stating that each symbol is in the ACL2 package, one can
  use extended package prefix syntax (if supported by your Lisp) to
  explicitly set the package for all symbols in a quoted S-expression,
  or import symbols from the ACL2 package as necessary.
\item \textbf{Be aware of ACL2's modified readtable} - ACL2 modifies
  the readtable of the Lisp that is hosting it. The most commonly
  problematic modification is the limitations that ACL2 places on the
  \code{\#.} read-time evaluation macro. To use certain libraries, you
  may need to temporarily restore the host's original readtable
  \code{*host-readtable*} before loading them.
\item \textbf{Use \queryfn\ for non-world-modifying queries, and
    \eventfn\ for world-modifying queries} - Though this is not
  currently enforced, we want to enforce it in the future, and at a
  minimum this provides useful documentation as to whether the author
  intended for a particular query to change ACL2's logical world.
\end{itemize}

\section{Specification}
\label{sec:specification}
Here, we provide a specification for the core functionality of the
three ACL2s interface functions. We first provide some definitions,
before going over the specifications for the \computefn, \queryfn\ and
\eventfn\ functions. We then review additional error handling cases.

This specification does not describe our output handling
functionality, which is intended to have no effect on the evaluation
of ACL2 forms other than to either suppress or capture any printed
output.

\subsection*{Definitions}
\begin{itemize}
\item An \textbf{ACL2 uterm} is a non-circular S-expression $x$ such
  that the ACL2 REPL's translation of $x$ satisfies the ACL2 predicate
  \code{termp}. This roughly corresponds to the notion of a
  syntactically valid ACL2 form. 

\item The \textbf{output signature set} $sig(x)$ of an ACL2 uterm $x$
  is a set $S$ containing lists of length 1 or greater (\textbf{output
    signatures}), where:
  \begin{itemize}
  \item $\code{(nil)} \in S$ iff there exist conditions under which
    $x$ evaluates to a single non-\stobj\ value.
  \item $\code{($j$)} \in S$ iff there exist conditions under
    which $x$ evaluates to a single value which is exactly the \stobj\
    $j$.
  \item $\code{($r_1\ r_2\ \cdots \ r_n$)} \in S$ iff there exist
    conditions under which $x$ evaluates to a multiple value of arity
    $n$ \code{($v_1\ v_2\ \cdots \ v_n$)} such that for all $i$,
    $r_i$ is \code{nil} iff $v_i$ is a non-\stobj\ value and $r_i$ is
    $j$ if $v_i$ is exactly the \stobj\ $j$.
  \end{itemize}
  This is inspired by the output specification of a constrained ACL2
  function's signature \seexdoc{SIGNATURE}. Note that, unlike ACL2
  functions, ACL2 terms are not required to have a single output
  signature. An arbitrary ACL2 term may contain conditionals where
  each branch of the conditional evaluates to a different output
  signature: for example, the ACL2 uterm \code{(if (boundp-global foo
    state) (@ foo) (mv 1 2 state))} has the output signature set
  $\{\code{(nil)}, \code{(nil nil state)}\}$ and the value that the
  term evaluates to may either be single value or an error triple
  depending on whether \code{state} has a bound global \code{foo}.

\item The \textbf{error triple} is the output signature \code{(nil
    nil state)}. This output signature is commonly encountered when
  using ACL2 functions that need to modify state, including all of the
  ACL2 event functions, so we provide special support for it over
  arbitrary multiple-value output signatures.

\item A term $x$ such that $sig(x)$ contains the error triple is
  considered to have encountered a \emph{soft error} during evaluation when
  it evaluates to an error triple such that the value in the first
  position of the error triple is \code{t}. This is consistent with the
  behavior of \code{(er soft ...)} \seexdoc{ER}.
\end{itemize}

\subsection*{Specification for \computefn}
\computefn\ takes a single argument $x$, which is expected to be an
ACL2 uterm such that $sig(x)=\{\code{(nil)}\}$. That is, $x$ is
expected to be a form that always evaluates to a single non-\stobj\
value. 

The return value of \computefn\ called on $x$ is defined as follows:
\begin{itemize}
\item if $x$ evaluates to multiple values or to a single value that is
  a \stobj, \computefn\ returns \code{(list t nil)}.
\item if the evaluation of $x$ results in a hard error, \computefn\
  returns \code{(list t nil)}.
\item if $x$ evaluates to a single value $v$ that is not a \stobj\ and
  no hard error occurred during evaluation, \computefn\ returns
  \code{(list nil $v$)}.
\end{itemize}

Note that the evaluation of $x$ can only result in a soft error if $x$
evaluates to an error triple, which is a multiple-value object. In
this case, the result of \computefn\ is specified by the first case
above.

\subsection*{Specification for \queryfn}
\queryfn\ takes a single argument $x$, which is expected to be an ACL2
uterm such that $sig(x)=\{\code{(nil nil
  state)}\}$. The evaluation of $x$ is expected to result in no calls
to any ACL2 event functions.

The return value of \queryfn\ called on $x$ is defined as follows:
\begin{itemize}
\item if $x$ evaluates to a single value or a non-error-triple
  multiple-value, \queryfn\ returns \code{(list t nil)}.
\item if the evaluation of $x$ results in either a soft or hard error,
  \queryfn\ returns \code{(list t nil)}.
\item if $x$ evaluates to an error triple \code{(nil $v$ state)} and
  no hard error occurred during evaluation, \queryfn\ returns
  \code{(list nil $v$)}.
\end{itemize}

\subsection*{Specification for \eventfn}
\eventfn\ takes a single argument $x$, which is expected to be an ACL2
uterm such that $sig(x)=\{\code{(nil nil state)}\}$.

The return value of \eventfn\ called on $x$ is defined as follows:
\begin{itemize}
\item if $x$ evaluates to a single value or a non-error-triple
  multiple-value, \eventfn\ returns \code{(list t nil)}.
\item if the evaluation of $x$ results in either a soft or hard error,
  \eventfn\ returns \code{(list t nil)}.
\item if $x$ evaluates to an error triple \code{(nil $v$ state)} and
  no hard error occurred during evaluation, \eventfn\ returns
  \code{(list nil nil)}.
\end{itemize}

If evaluating $x$ led to a hard or soft error, ACL2's world is
``reverted'' to its value prior to the evaluation attempt. By
``revert,'' we mean ACL2's notion of reverting (as referred to in
e.g. the \code{ld-error-action} XDOC topic), which is subtle as some
values may not be reverted. \eventfn's behavior is consistent with
ACL2's notion of reversion.

\subsection*{Error Handling}
\begin{itemize}
\item If \computefn, \queryfn\ or \eventfn\ are called with an
  argument that is not an ACL2 uterm, each will return
  \code{\textquotesingle(t nil)}. Practically, our implementations of
  these functions delegate the responsibility of determining whether
  the term that the user provided is ``valid'' to \code{ld}.
\item If \queryfn\ is called on an ACL2 uterm that adds, removes, or
  modifies ACL2 events, then we do not make any claims as to whether
  or not the ACL2 world will be modified. 
\end{itemize}

When an ACL2s interface function reports that evaluation resulted in
an error, it may be difficult to determine the exact error that
occurred. This is because ACL2 does not make error causes easily
accessible through return values. In a pinch, one can use the output
capture functionality of the ACL2s interface and some text parsing to
attempt to identify the causes of errors.

\section{Applications}
\label{sec:applications}

We have used the ACL2s Systems Programming interface in several
applications. Our experience in doing so has helped inform the design
and functionality of the interface. We briefly discuss several of
these applications below.

\subsection*{Building a Theorem Prover}

The ACL2s interface was first used at scale in Northeastern's CS4820
\textit{Computer Aided Reasoning} course. In this course, students
build a first-order theorem prover in systems ACL2s over the course of
the semester. Being able to easily call ACL2s from Common Lisp is
useful throughout the process of building a theorem prover:
\begin{itemize}
\item One can verify that propositional simplification works correctly
  on a test suite by asking ACL2 to prove that the original statements
  are propositionally equivalent to the simplified statements. 
\item It is possible to support first-order atomic formulae that
  include terms that allow the use of ACL2s functions. Such terms and
  formulae can be validated by checking that the arity and types of
  arguments are consistent with the underlying ACL2s functions.
  Terms can also be simplified if they involve functions all of whose
  arguments are constants. 
\item Students and instructors use ACL2s' data type enumeration
  facilities to automatically generate terms, formulae and other types
  for testing purposes, after defining an appropriate \code{defdata}
  data type.
\end{itemize}

Meanwhile, the fact that the students are writing their theorem prover
in systems ACL2s means that they have access to general Common Lisp
functionality that would not be available even if using writing
program-mode core ACL2s functions, including higher-order functions,
external libraries, the powerful Common Lisp \code{loop} macro,
full Common Lisp argument handling (e.g. keyword and optional
arguments) and CLOS (the Common Lisp Object System). Performance is an
important consideration for the students when developing certain parts
of their theorem provers, and systems ACL2s allows the students to
both use familiar data structures like hash tables as well as
destructive operations on those data structures to write efficient
code without needing to learn how to program using \stobj s. When
using systems ACL2s, students are still free to write whatever code
they would like to reason about in core ACL2s and easily call it from
their systems ACL2s code. This flexibility allows students to decide
how they would like to partition their system between systems and core
ACL2s code.

\subsection*{Invariant Discovery Games}
Another area of research interest for the authors is the gamification
of loop invariant discovery. In our first work in this
area~\cite{idg}, we show that it is possible to develop a game that
allows programmers without formal methods expertise to assist an
automated tool in discovering loop invariants for the verification of
imperative programs, while displaying to the players the code that was
being reasoned about. Prior work had focused on games that did not
display code to players for a variety of reasons, including
requirements for code confidentiality, concern about high cognitive
load for players, and a desire to support players without programming
expertise. The games that we developed (IDG and IDG-T) did not
directly use the ACL2s interface, but did use a similar style of
interaction with ACL2 on the backend. We developed a wrapper around
ACL2 consisting of a large set of ACL2 macros and some Python code
that allowed us to interact with ACL2 from a browser application. The
end result is a system such that ACL2 is hidden completely from the
end-user.

As the ACL2s interface matured, we began to write more of our
invariant discovery game code in systems ACL2s. This is highlighted by
our later preprint~\cite{reasoning-engine-preprint} that discusses an
improved version of the backend used in IDG and IDG-T. This new
backend is described in a theorem-prover-agnostic manner, but relies
on the existence of an interface that allows one to send the theorem
prover a verification condition and receive back either a
counterexample showing that the verification condition does not hold,
a message indicating a proof was successful, or a message indicating
that no counterexample could be found but the verification condition
also could not be proved. This interface is easily implemented using
ACL2s and the ACL2s interface.

\subsection*{Autograding}

We have used the interface to develop autograders for Northeastern's
\textit{Logic and Computation} course, where students learn how to
reason about computation alongside ACL2s. Combining ACL2s'
property-based testing functionality with the interface allows us to
quickly develop executables that use external libraries to, for
example, output JSON for ingestion by a learning management system's
(LMS)'s autograder functionality. In this way, we can provide students
with low-latency feedback on their submissions, in many cases
providing them with concrete examples that highlight how their
function definitions do not fully satisfy the specification we
provided them with.

\subsection*{Proof Checker}
We have developed a more complex autograder---a proof checker---for
equational reasoning proofs written in a format appropriate for an
introductory course like \textit{Logic and Computation}. This system
interfaces with several external tools when it processes input, and as
such contains a significant amount of raw Lisp I/O code that would be
difficult to reason about. The proof checker also makes liberal use of
Common Lisp's condition system to pass information about errors across
multiple levels of functions, attaching information useful for
localizing the source text responsible for the error along the
way. Writing an analogous system using core ACL2 would require that
every function, even those that are intended to simply pass errors
through, be aware of which functions it calls may return errors, so it
can pass them through. Adding an error to a function in such a system
may therefore require changes to several other functions, since they
must now either explicitly handle that error or explicitly pass it
through. The condition system provides some additional functionality
that may be difficult or impossible to emulate in core ACL2, like
restarts that allow top-level functions to control how to handle
errors in deeply nested functions.

\subsection*{Fuzzing}
The authors have been involved in work regarding the use of ACL2s'
data definition and counterexample generation facilities to fuzz
complicated protocols that may have many relationships within and
between messages. We developed a proof-of-concept fuzzer for the PTaaS
(path tracing as a service) service that was included in DARPA's CGC
(Cyber Grand Challenge) competition. The fuzzer was implemented in
systems ACL2s using the ACL2s interface, and communicated with a test
harness that monitored and managed the PTaaS process as it was being
fuzzed. In this application, we mainly made direct calls from Common
Lisp to the ACL2s enumerators for the messages we wanted to generate,
though there were a few calls (\eg\ to set the ACL2s RNG seed) that
needed to be made through the interface functions. The performance
requirements of this application meant that it would likely be
necessary for the systems ACL2s code to directly communicate with the
harness (\eg\ over a TCP socket), which was straightforward given the
\code{usocket} Quicklisp library~\cite{usocket}.

As part of the fuzzing work, we developed a library for managing pools
of ACL2s processes and communicating with them over TCP using a simple
protocol. By using TCP sockets to communicate with the ACL2s
processes, our library has the flexibility to support systems that
distribute ACL2s processes over a number of computers connected to the
same network, not just systems that run multiple ACL2s processes on a
single machine.

\subsection*{Z3 Interface}
Another benefit of using the ACL2s interface is that it allows one to
use ACL2 in conjunction with additional external tools. We have
developed an interface that allows one to interact with the Z3 SMT
solver from Common Lisp (and thus also from systems ACL2s code). The
syntax of queries was intentionally chosen to mirror the syntax of
ACL2s interface queries. Thus, when students in the \emph{Logic and
  Computation} course used the Z3 interface to develop simple Sudoku
solvers, they could adapt much of the knowledge they had already
developed from using the ACL2s interface previously. This ability to
provide similar or identical interfaces for various solver backends is
powerful, and is something that we believe is much easier to do using
systems ACL2s rather than core or systems ACL2.

We note that our Z3 interface differs from Peng \etal's Smtlink
interface~\cite{peng-smtlink}~\cite{smtlink2} insofar as our interface does
not seek to translate ACL2 expressions into Z3 formulae. Instead, our
interface allows a user to directly generate Z3 assertions and ask Z3
for models that satisfy the assertions. Since the user must write the
code to generate Z3 assertions themself, our interface does not
provide the ability that Smtlink does to add a hint to ACL2 and get
SMT-generated proofs or counterexamples without extra work. On the
other hand, our Z3 interface is able to support many more Z3 features
than Smtlink does, including custom Z3 sorts, regular expressions,
optimization, function and array values, quantifiers, sets, and more.

\section{Related Work}
\label{sec:related}
ACL2's core depends on ``raw Lisp,'' since it needs to bootstrap
itself inside of a Common Lisp environment. ACL2 also provides ``raw
Lisp'' functionality that can be accessed through the use of trust
tags or by simply exiting the ACL2 REPL.  Our work is built on top of
this ``raw Lisp'' functionality.

The ACL2 Sedan programming environment~\cite{dillinger-acl2-sedan} and
the ACL2 Bridge~\cite{davis-acl2-bridge} are two projects that allow
ACL2 to interoperate with another programming language. The ACL2 Sedan
makes several modifications to the ACL2 REPL that allow it to extract
more information about the result of an evaluation, and operates by
running an ACL2 subprocess inside of the host Java program, hooking
the subprocess' standard input and output streams up to input and
output streams that the host Java program controls. The ACL2 Bridge
presents the ACL2 REPL over either a UNIX or a TCP socket, with
support for formatting results of a request as either S-expressions or
as JSON objects. Both systems make use of trust tags, as they both
need to perform potentially unsound operations to function properly
(the ACL2 Sedan overwrites some built-in ACL2 functions so that it can
extract more information from ACL2; the ACL2 Bridge contains a
significant amount of ``raw Lisp'' code for interacting with sockets
and redirecting ACL2's output). The methodology described in this
paper does not directly provide interoperation with programming
languages running outside of ACL2's host Lisp, but it is useful when
developing a system that does need to perform such interoperation.

Other theorem provers, proof assistants, and SAT/SMT solvers provide
external APIs that allow users programmatic control over the
underlying software. As examples, Z3 provides core C and C++
interfaces~\cite{z3} and Isabelle provides a Scala
interface~\cite{async-proof-processing}.

\section{Conclusion}
\label{sec:conclusion}
In this paper, we presented the ACL2s systems programming methodology
and the ACL2s interface functions at its core. We described our
experiences using the ACL2s systems programming methodology in several
research projects. We hope that the ACL2s interface library will be
useful to others who develop tools using ACL2 and ACL2s as key
components.

\subsection*{Acknowledgments}
We thank everyone who worked with us on projects that used the ACL2s
system programming methodology, including Ben Boskin, Seth Cooper,
Dave Greve, Ankit Kumar, Benjamin Quiring and Atharva Shukla, as well
as the students of Northeastern's CS4820 and CS2800. Their feedback
has been invaluable. We thank the reviewers of this paper. Their
feedback helped us improve both this paper and the ACL2s interface
library. Finally we thank J Strother Moore, Matt Kaufmann, and the
ACL2 community for the design, implementation and continued
maintenance of the ACL2 system and its libraries.

\nocite{*}
\bibliographystyle{eptcs}
\bibliography{paper}
\end{document}